**LDx: estimation of linkage disequilibrium from high-throughput pooled resequencing data**

Alison F. Feder[1,2*], Dmitri A. Petrov[1] and Alan O. Bergland[1]

[1]Department of Biology, Stanford University, Stanford, USA

[2]Department of Mathematics, University of Pennsylvania, Philadelphia, USA

*afeder@stanford.edu

## ABSTRACT

High-throughput pooled resequencing offers significant potential for whole genome population sequencing. However, its main drawback is the loss of haplotype information. In order to regain some of this information, we present LDx, a computational tool for estimating linkage disequilibrium (LD) from pooled resequencing data. LDx uses an approximate maximum likelihood approach to estimate LD ($r^2$) between pairs of SNPs that can be observed within and among single reads. LDx also reports $r^2$ estimates derived solely from observed genotype counts. We demonstrate that the LDx estimates are highly correlated with $r^2$ estimated from individually resequenced strains. We discuss the performance of LDx using more stringent quality conditions and infer via simulation the degree to which performance can improve based on read depth. Finally we demonstrate two possible uses of LDx with real and simulated pooled resequencing data. First, we use LDx to infer genomewide patterns of decay of LD with physical distance in *D. melanogaster* population resequencing data. Second, we demonstrate that $r^2$ estimates from LDx are capable of distinguishing alternative demographic models representing plausible demographic histories of *D. melanogaster*.

## 1. INTRODUCTION

Linkage disequilibrium (LD) is a measure of the association between alleles at two loci encapsulating how often these alleles are observed together. LD is an important statistic because it reflects the historical rates of recombination between loci and thus forms the basis for many tests of selection [1] and the estimation of demography [2,3]. Measurement of LD fundamentally requires knowledge of multi-locus haplotype frequencies within a species and these frequencies have been traditionally obtained through direct observation of haplotypes or statistical inference of haplotypes from unphased genotype data [4,5]. While these approaches are feasible for single locus studies, they can become logistically and computationally difficult when applied genomewide.

Here, we present a simple and cost effective method to directly measure short-scale LD genomewide using pooled next-generation resequencing data without any prior knowledge of genotype frequencies or of the haplotypes present in the population. Pooled resequencing data is generated by anonymously mixing DNA from multiple individuals from a population or species followed by massively parallel sequencing. Pooled resequencing occurs naturally when sequencing intrinsically heterogeneous samples (e.g., tissue samples from one individual or microbe communities) and is becoming a common experimental technique for quantitative [6] and population genetic [7,8] analyses. Pooled resequencing is a highly accurate method to estimate SNP [9-16] frequencies and has also been used to estimate haplotype frequencies from pooled samples when haplotypes are known *a priori* [17].

While there is some debate concerning the use of pooled resequencing versus simply sequencing strains individually, both methods have merits in different situations and certain scenarios necessitate or benefit from the use of pooled sequencing (see Futschik and Schlotterer [18] and Cutler and Jensen [19] for extensive discussion). For instance, in some cases individual genomes cannot be isolated (e.g., tissue samples). In other circumstances, often encountered in evolutionary applications, sampling many individuals of a population is easy but sequencing them is labor-intensive or prohibitively expensive. Although pooled resequencing has proved useful in measuring allele frequencies to assess population differentiation [20] and summary statistics based on the site frequency spectrum, [21] researchers often forfeit estimates of linkage between polymorphic loci because of the limited haplotype information available in an experiment utilizing pooled resequencing.

---

*To whom correspondence should be addressed.

We demonstrate that some of this haplotype information can be reclaimed on a short scale which nonetheless allows genomewide patterns of linkage to be observed. Our approach, called LDx, directly estimates LD from pooled samples by measuring two-locus haplotype frequencies across short sequence reads that tile any particular genomic region. We test the accuracy of our technique empirically by estimating $r^2$, a common measure of LD, in a pooled sample of 92 wild type *Drosophila melanogaster* with individually sequenced genomes [22]. We find that our technique accurately estimates $r^2$ across the genome and that the correlation between the pooled and actual estimates of $r^2$ is in the expected range given the sampling variance determined by the read depth of our samples. Finally, we show two applications of LDx: first, we demonstrate that estimates of $r^2$ based on pooled samples show a classic signature of decay with physical distance and that the rate of decay is negatively correlated with recombination rate; second, we use LDx to investigate two alternative demographic histories of *D. melanogaster*. LDx is implemented as an open-source Perl script available via sourceforge (https://sourceforge.net/projects/ldx/ ).

# 1 METHODS

**Calculation of haplotype tables.** To generate two-locus haplotype tables, LDx takes a list of sites that are polymorphic within the pooled sample and a file containing the positional mapping information of each read, specified in the SAM format [23]. The position of polymorphic sites can be inferred directly from the pooled sequence data using a variety of techniques [24] or can be a list of polymorphisms known *a priori*. LDx then finds all reads that cover pairs of polymorphic sites whose distance apart is less than the maximum insert size of the sequencing library. As is shown in Figure 1, the count for each two locus haplotypes is computed, where $x_{ij}$ is the number of genotypes observed with allele $i$ at the first locus and allele $j$ at the second locus. We refer to the number of reads that cover both polymorphic sites as the *intersecting read depth*. $r^2$ is calculated between pairs of sites with intersecting read depth greater than a minimum threshold, by default ten. In the case of loci with more than two alleles, LDx takes the two most frequent alleles and reports $r^2$ estimates with reference to those.

**Method 1 – direct inference.** LDx reports the $r^2$ value that would be calculated by naive observation of the haplotype table. That is, it is computed as

$$r^2 = \frac{(\frac{x_{AB}}{x_{AB}+x_{Ab}+x_{aB}+x_{ab}} - p_A p_B)^2}{p_A p_B (1-p_A)(1-p_B)}$$

where $p_A$ and $p_B$ are the allele frequencies computed only from the intersecting reads:

$$p_A = \frac{x_{AB}+x_{Ab}}{x_{AB}+x_{Ab}+x_{aB}+x_{ab}} \qquad p_B = \frac{x_{AB}+x_{aB}}{x_{AB}+x_{Ab}+x_{aB}+x_{ab}}$$

**Method 2 – approximate maximum likelihood.** To estimate $r^2$ using maximum likelihood, LDx uses the observed haplotype table and allele frequency estimates derived from all reads covering the two loci. We estimate allele frequencies $p_A$' and $p_B$' using total read depth rather than from the marginal allele frequencies calculated from the haplotype table because estimates made from all reads will be more accurate than estimates just made from intersecting reads.

For each pair of sites, we estimate $r^2$ by computing the maximally likely $r^2$ conditional on the observed allele frequencies. While the observed frequencies represent only an approximation to the true frequencies, they act as a useful proxy for the purpose of evaluating the likelihood of $r^2$. The likelihood of the observed haplotype table conditional on $r^2$ and the observed allele frequencies is,

$$P(x_{AB}, x_{Ab}, x_{aB}, x_{ab} | p'_a, p'_b, r^2) = \frac{(x_{AB} + x_{Ab} + x_{aB} + x_{ab})!}{x_{AB}! x_{Ab}! x_{aB}! x_{ab}!} \hat{f}_{AB}^{x_{AB}} \hat{f}_{Ab}^{x_{Ab}} \hat{f}_{aB}^{x_{aB}} \hat{f}_{ab}^{x_{ab}}$$

where $f_{ij}$ is the expected proportion of haplotype $ij$ given $r^2$ and is computed

$$\hat{f}_{AB} = \sqrt{r^2 p'_A (1-p'_A) p'_B (1-p'_B)} + p'_A p'_B$$
$$\hat{f}_{Ab} = \sqrt{r^2 p'_A (1-p'_A) p'_B (1-p'_B)} - p'_A (1-p'_B)$$
$$\hat{f}_{aB} = \sqrt{r^2 p'_A (1-p'_A) p'_B (1-p'_B)} - (1-p'_A) p'_B$$
$$\hat{f}_{ab} = \sqrt{r^2 p'_A (1-p'_A) p'_B (1-p'_B)} + (1-p'_A)(1-p'_B)$$

and the allele frequencies are estimated as follows:

$$p'_A = \frac{x_{Ab} + x_{AB} + x_{A-}}{x_{AB} + x_{Ab} + x_{aB} + x_{ab} + x_{A-} + x_{a-}}$$

$$p'_B = \frac{x_{Ab} + x_{AB} + x_{-B}}{x_{AB} + x_{Ab} + x_{aB} + x_{ab} + x_{-B} + x_{-b}}$$

Using this approach, the most likely linkage disequilibrium estimate can only be computed for SNP pairs where the allele frequencies estimated across all reads are congruent with the haplotype table estimated from the intersecting reads. Because the intersecting reads are a subset of the total number of reads, such incongruent estimates are likely to occur when the true $r^2$ is high, but not equal to one. When allele frequency estimates are incongruent with the haplotype table, the maximum likelihood is undefined and the reported maximum likelihood is at the boundary of the likelihood surface. LDx reports information on whether the $r^2$ estimate for a particular pair of sites is likely to be undefined.

This method is labeled as approximate, because it assumes the observed allele frequencies as true, instead of simultaneously maximizing the probabilities of the observed allele frequencies and $r^2$. Our implementations of the simultaneous three variable maximization frequently failed to converge. In scenarios in which our estimates did converge, the true MLE and approximate MLE yielded similar results (results not shown). We therefore report the approximate MLE $r^2$ as a faster and more reliable proxy estimate.

**Accounting for the experimental design.** In pooled resequencing experiments, the binomial (multinomial) variance associated with esimates of allele (haplotype) frequency are a function of the number of chromosomes sampled and the number of reads at any locus (supplemental equation 3 in [25]). The variance of frequency estimates can be easily approximated by calculating the effective number of observations at a given locus, conditional on read depth and number of chromosomes in the sample, as

$$n.eff = \frac{n.reads * n.chr - 1}{n.reads + n.chr}$$

We use this formula to calculate the effective number of observations for each two-locus genotype when calculating the approximate maximum likelihood estimates of $r^2$ above. LDx uses the effective number of observations to estimate the 95% confidence intervals surrounding the MLE estimate. Confidence intervals are calculated as ±1.96 log-likelihood units away from the MLE (see the users guide).

**Empirical validation.** To test the accuracy of $r^2$ estimation from pooled resequencing, we used short read data described elsewhere ([16] SRA accession SRR353365.1). Briefly, this library is a pool of 92 highly inbred *D. melanogaster* strains derived from a natural population in Raleigh, North Carolina representing a subset of the 162 strain Drosophila Genetic Reference Panel (DGRP, [22]). Average autosomal coverage in this library is ~40X and average coverage of the X-chromosome is ~20X. Only reads with base quality scores > 20 were used. We identified all biallelic SNPs in the DGRP population that are fixed within each strain (i.e., sites with no residual heterozygosity) using precomputed SNP tables (https://www.hgsc.bcm.edu/content/drosophila-genetic-reference-panel). Of those, we only considered sites in which the total read depth in the pooled sample was less than twice the chromosomal average in order to exclude potential copy number variants from the analysis. Our analysis also includes investigation of the accuracy of $r^2$ estimates based on the number of intersecting reads and the observed minor allele frequency.

**Simulation.** To test whether the observed correction between $r^2$ estimated from pooled data and the DGRP is expected given binomial sampling, we generated simulated reads from the DGRP data. To generate simulated pooled paired end reads, we used *wgsim* (23). *wgsim* accepts a FASTA file listing full haplotypes from multiple individuals and simulates the pooling process as if sampling from a population composed of these individuals at user specified read depths, read lengths and gap sizes. We used *wgsim* to simulate a population composed of the 85 DGRP strains with 93bp paired end reads at ~10x, 40x, 100x and 200x coverage. Note, we simulated a pooled population of 85 that are a perfect subset of the 92 strains used in the experimental pooled resequencing study; we were unable to simulate pooled resequencing for all 92 because 7 strains were not sequenced to sufficiently high coverage.

We generated estimates of $r^2$ from these libraries as described above, with a minor allele frequency cutoff of 1%, and a minimum intersecting read depth of 10, except that for the simulated 10X library, in which we only required a minimum of 5 intersecting reads.

## 2 RESULTS

LDx represents, to our knowledge, the first effort to estimate levels of linkage disequilbrium from pooled resequencing data directly with no prior information of haplotype frequencies. The one existing method to infer haplotype frequencies and levels of LD from pooled data [17] requires prior knowledge of haplotype frequencies in the population. Obtaining prior knowledge of genomic haplotypes can be difficult, expensive and labor intensive. Moreover, the method presented in Long *et al*. [17] as well as analogous methods to phase di- and polyploid sequence data (e.g., [4,26]) likely perform best when prior haplotypes are drawn directly from the population in question. This requirement limits the utility of these approaches. Through bypassing the haplotyping step, LDx can be applied to populations for which only pooled resequencing data exist.

**Two-locus haplotype reconstruction:**

LDx recovers sufficient data from the pooled paired end resequencing data to make inferences of linkage disequilibrium through identifying SNP pairs with many intersecting reads. LDx is able to detect SNP pairs that fall both on a single read and across paired end reads, creating a bimodal distribution on the distances between two SNPs of an identified SNP pair (Figure 2A). As read depth increases in our simulations, we find that the proportion of SNP pairs where $r^2$ can be estimated by the approximate maximum likelihood methods increases (Figure 2B).

**Empirical validation:**

$r^2$ estimates from pooled samples were highly correlated with estimates from the actual haplotype data (*p*-values for all correlation coefficients << 0.001, Figure 3AB). For the *direct estimation* method, we observed a small amount of upward bias in our observed estimates of $r^2$ due to sparse sampling of the haplotype tables, leading to $r^2$ estimates at 1. This upwards bias was not present in the *MLE* method since estimates integrated both the allele frequencies and the observed haplotype tables. We observed a small amount of downward bias in our *MLE* estimates, because incongruities between allele frequency estimates and observed haplotype frequencies caused $r^2$ estimates of zero when only a subset of the haplotype table was sampled. The accuracy of $r^2$ estimates by LDx increases with higher minor allele frequency (Figure 3D). $r^2$ is more accurately estimated for these pairs because there is a high probability of observing all possible haplotypes.

**Dependence on read depth, read length and insert size:**

In simulations of different read depths, we found that increasing read depth leads to an increase in the correlation between DGRP $r^2$ and $r^2$ estimated by the *direct estimation* and *MLE* methods (Figure 3C). The observed correlation estimate between the DGRP $r^2$ and both *direct estimation* and *MLE* $r^2$ from the NC92 data fell within the range of correlation estimates produced by our simulations. This serves as a validation of our simulation procedure.

     Given these results, increasing read length (and keeping the number of reads constant) is expected to increase the accuracy of $r^2$ estimates because read depth at any given locus will increase (results not shown). However, increasing insert size will generally decrease the accuracy of $r^2$ estimates because the average intersecting read depth for any two SNP pairs will be lower. To see this, note that the variance of insert size scales proportionally to the average insert size. Thus, increasing the insert size will decrease the intersecting read depth particularly for pairs of SNPs that are at the average distance between the paired end reads.

**Decay of LD with distance:**

To test that estimates of $r^2$ made by LDx are biologically meaningful, we measured the decay of $r^2$ with physical distance in our pooled resequencing data. LDx estimates of $r^2$ show the classic pattern of decay with physical distance (Figure 4) and the rate of decay varies as a function of recombination rate in a pattern highly congruent with the decay rate of true $r^2$ estimates (Table 1). In regions of low recombination, the rate of decay of LD is higher than in regions of high recombination. This is because at very short physical distance (e.g., less than approximately 100bp), loci in regions of low recombination are highly linked (high $r^2$) whereas loci in regions of high recombination are less tightly linked (lower $r^2$). However, by ~300 bp, loci in regions of both low and high recombination have similar patterns of linkage.

**Use of LDx in differentiating between demographic events**

Estimates of the site frequency spectrum and their deviation from the expectation under neutrality can be useful for identifying demographic events [27]. However, in some situations, alternative demographic events can result in populations with very similar levels of polymorphism. For instance, following a population bottleneck we expect a reduction in heterzygosity that is proportional to the the duration and the magnitude of the bottleneck. To see this, note that expected heterozygosity following a bottleneck can be computed as,

$$H_t = H_0 \, e^{-\frac{t}{N_b}}$$

[28], where $H_t$ is the post-bottleneck estimate of heterozygosity, $H_0$ is the initial heterozygosity, t is the duration of the bottleneck and $N_b$ is the size of the bottleneck population. Therefore, a population with a bottleneck half as severe but with a duration twice as long as some original population will have an identical estimate of heterozygosity, measured as π. However, the LD between sites in these two populations may not necessarily the same. In these situations, LDx can be used to distinguish these models.

We measured π using Variscan [29] and $r^2$ in a forward-simulated population run in SFS_code [30] for an out of Africa bottleneck in *D. melanogaster* [31] (see figure 5). We then repeated the simulations in two additional simulated populations – one with a bottleneck twice as large, but lasting half as long (severe), and one with a bottleneck half as large but twice as long in duration (mild). The average $r^2$/bp estimated both by the approximate MLE method and the direct computation are reported in table 2.

LDx estimated a significantly higher average $r^2$/bp for both the approximate MLE and direct estimation $r^2$ values for the severe model when compared to the original model (p-values 0.014 and 0.007, respectively). While LDx did not report a significantly lower $r^2$/bp for the mild bottleneck model, it was significantly different from the severe model (p-values 0.0013 and 0.0012, respectively).

## 3 DISCUSSION

LDx represents, to our knowledge, the first effort to directly estimate levels of linkage disequilibrium from high-throughput pooled resequencing data with no prior knowledge of haplotype structure in the target population. It provides an accurate estimate of linkage over hundreds of basepairs genomewide, and suggests that important information on linkage can be retrieved from populations sequenced using pooled sequencing. Note, however, that our ability to estimate LD accurately between any two specific points is low even at reasonably high sequencing depths and even if they are physically close to each other, because the number of reads that overlap any two particular SNPs is much lower than the coverage at any one specific SNP (Fig. 2B).

Certain conditions make the extraction of useful LD information from pooled data very difficult. For example, if the read length of the pooled sequences is much shorter than the length at which linkage decays to background levels in the genome, LDx will not provide informative output concerning $r^2$. Further, linkage cannot be calculated beyond the length of a read pair, as haplotyping is impossible with pooled data. Indeed, those researchers interested in identifying faint signals at long distances may have better success with individual strain haplotyping. Additionally, if genomic polymorphisms are very sparse, LDx will estimate linkage based on a small number of pairs. Such limitations make it unlikely that LDx or similar methods will useful for humans or other organisms with low levels of polymorphism per basepair.

Despite these limitations, we imagine estimates of $r^2$ made by LDx will be useful in understanding how patterns of LD change genomewide due to selection and demography. For instance, strong bottlenecks are expected to dramatically increase pairwise LD genomewide and the average change in LD before and after a bottleneck could be used to estimate the severity of the bottleneck [32]. As demonstrated above, certain disparate demographic effects will leave similar imprints in the site frequency spectrum. LDx offers the potential to differentiate these scenarios by detecting differences in linkage. LDx could also be useful for identifying previously unannotated paralogs as these regions should have aberrantly high estimates of LD.

As sequencing technology continues to improve, read depth and fragment length will increase. This will result in a higher accuracy of $r^2$ estimation and an increase in the probability that $r^2$ can be estimated between two SNPs. While these improvements will only marginally increase the accuracy of allele frequency estimation, they will dramatically increase the accuracy of LD estimation from pooled data.

**Figure 1. Cartoon depicting information leveraged from pooled paired end reads.** The cartoon represents an example observation between two loci. Although many reads hit one locus or the other, only five reads cross both loci. In this example, $p_A$, computed only from intersecting reads, is 3/5, while $p_A$', computed from all available reads is 4/8.

**Figure 2. Identification of SNP pairs.** A) The distance between component SNPs of a SNP pair are bimodally distributed, reflecting the frequency of pairs that fall within a single read or across paired end reads. B) Increasing the read depth increased the proportion of pairs it was possible to locate in the pooled paired-end read data with a 0.01 allele frequency cutoff. This proportion of estimable pairs is calculated by counting the number of SNPs in a moving window of length 300 bp and using that to compute the number of possible SNP pairings (*n* choose 2). This is then compared to the number of SNP pairs identified at a given read depth.

**Figure 3. Method performance of LDx in predicting linkage.** $r^2$ measured from the DGRP haplotypes is strongly correlated with estimates from A) the *direct observation* method and B) the *maximum likelihood* method. In A), observing only a sparse sampling of the haplotypes creates the overabundance of observed $r^2$ estimates of 1. We determined the correlation between our $r^2$ estimates and $r^2$ values derived from haplotype data provided by the DGRP (Mackay *et al* 2012). We restricted the DGRP dataset to those strains present within our sample (92 of 162 strains). C) Increasing the simulated read depth increased the correlation between the true $r^2$ and the $r^2$ estimated by the direct observation (red) and maximum likelihood (blue) methods. Estimates in these figures have minor allele frequency cutoff of 1%. **D)** Filtering based on minor allele frequency leads to more accurate $r^2$ estimates for the *direct observation* (red) and *maximum likelihood* (blue) methods. Points represent $r^2$ estimates made from pooled resequencing of the DGRP.

**Figure 4: LDx predictions decay at a biologically plausible rate.** $r^2$ decays in a similar pattern among the *direct estimation* (red), *maximum likelihood* (blue) and DGRP (green) $r^2$ measures. Points represent average $r^2$ within distance classes. Averages were applied only to pairs that had minor allele frequency > 0.1. Lines represent predicted decay or $r^2$ with physical distance. Decay models were fit in R 2.13 (R core Development Team 2012).

**Figure 5: Reference Demographic Model.** Following Table 2 in Thornton & Andolfatto's out of Africa model [31] at $\varrho/\theta = 7$, the population reaches equilbrium at population size $N_0$, contracts to a size of $N_b$, and then expands back to $N_0$ after $4N_0 t$ generations. The population then continues another $4N_0$ (.048) generations before sampling. In our model, we used $N_0 = 1000$ and sampled 20 individuals.

**Table 1.** Comparison of the decay of $r^2$ with distance and recombination rate as estimated by different methods.

| Parameter | True $r^2$ | *Direct observation* $r^2$ | *Approx. maximum likelihood* $r^2$ |
|---|---|---|---|
| **Intercept** | 0.662 ± 0.008 | 0.654 ± 0.009 | 0.609 ± 0.007 |

|  | (82.51) | (75.729) | (87.97) |
|---|---|---|---|
| **log(distance)** | -0.0757 ± 0.0007 (-104.4) | -0.0604 ± 0.0008 (-77.37) | -0.0729 ± 0.0006 (-116.6) |
| **recombination rate** | -0.022 ± 0.0012 (-18.07) | -0.018 ± 0.0013 (-13.63) | -0.017 ± 0.001 (-16.52) |
| **log(dist) x rec. rate** | 0.00609 ± 0.0007 (8.50) | 0.00661 ± 0.0008 (8.575) | 0.00626 ± 0.0006 (10.14) |

Results from a regression model that examines the how $r^2$ decays as a function of physical distance (bp) and recombination rate (cm/Mb) and their interaction. Recombination rates were estimated from Fiston-Lavier *et al*. (2010) [33]. Values represent parameter estimates ± standard error and t statistics (in parentheses).

**Table 2.** Comparison of $r^2$ values in population with bottlenecks producing similar average pairwise differences (π).

| Model | N0/Nb | t[a] | π | App MLE $r^2$/bp | Direct Est. $r^2$/bp |
|---|---|---|---|---|---|
| **Thornton & Andolfatto [31]** | 0.047 | 0.021 | 0.0029 | 0.011±0.078 | 0.0125±0.082 |
| **Severe Bottleneck** | 0.0235 | 0.042 | 0.0028 | 0.005±0.035 | 0.0056±0.036 |
| **Mild Bottleneck** | 0.094 | 0.0105 | 0.0028 | 0.026±0.123 | 0.0266±0.122 |

Different models of demography can result in very different linkage patterns while retaining similar π values making it difficult to differentiate the models using frequency spectrum based methods. LDx can distinguish between the models by estimating $r^2$. The model with a mild, more prolonged bottleneck has lower $r^2$ than the Thornton & Andolfatto reference model, while the model with a more severe, short bottleneck has higher $r^2$. Estimation cells report mean $r^2$ ± standard deviation.

[a] t is measured in generations scaled by $4N_e$. All models had the same number of post bottleneck generations as the Thornton & Andolfatto model

Fig 1.

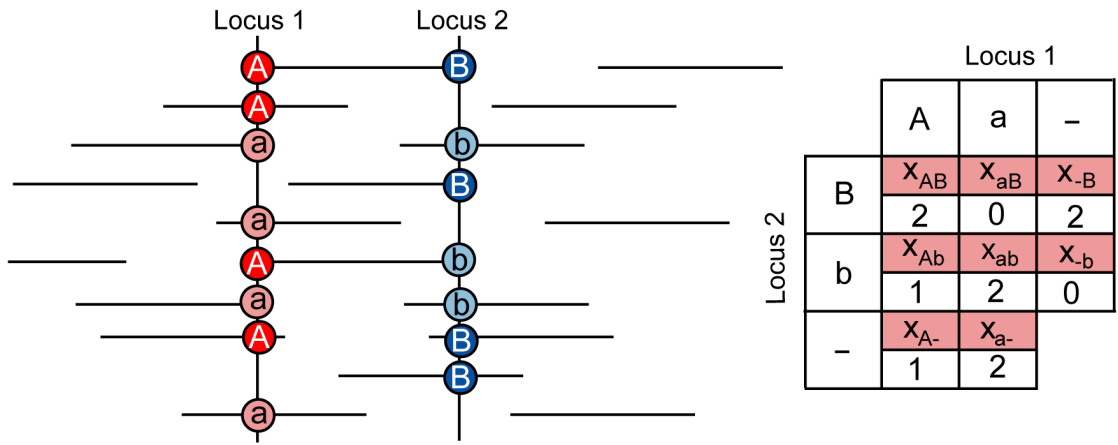

Fig. 2

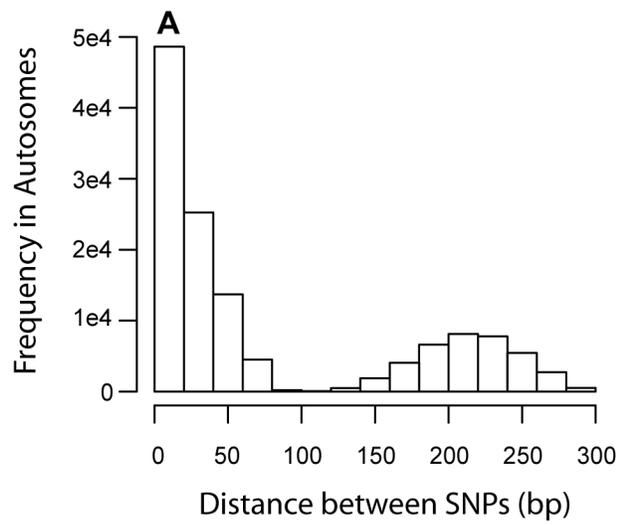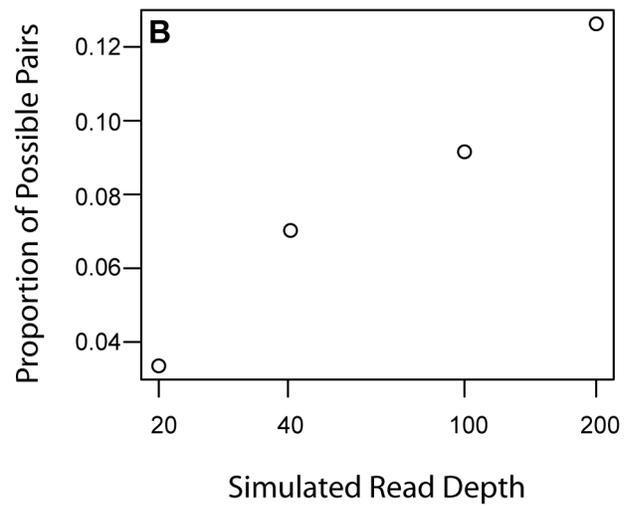

Fig. 3

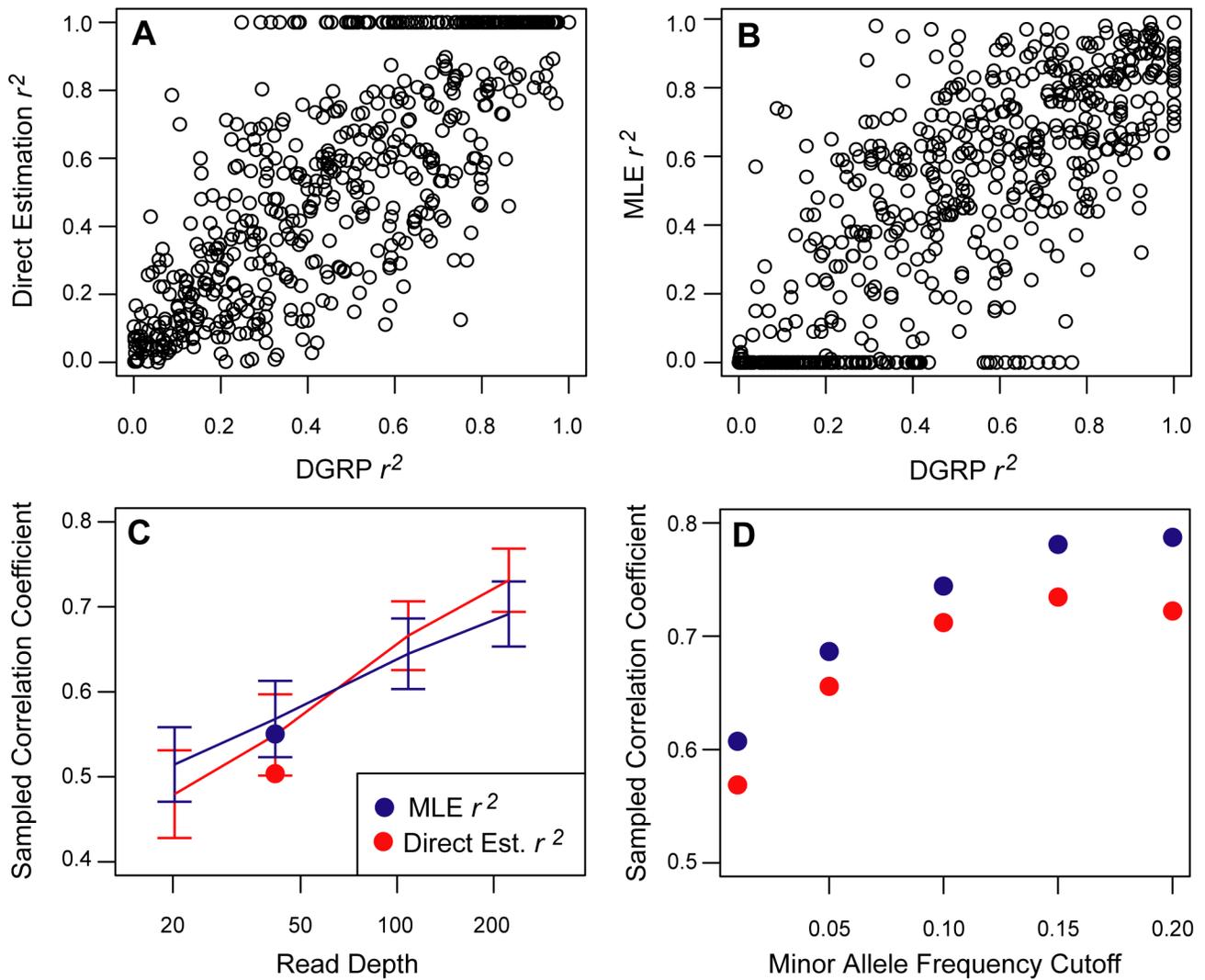

Fig 4

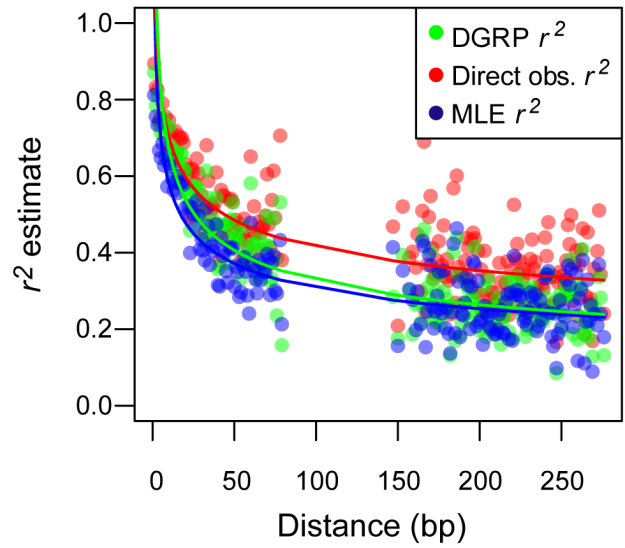

Fig. 5

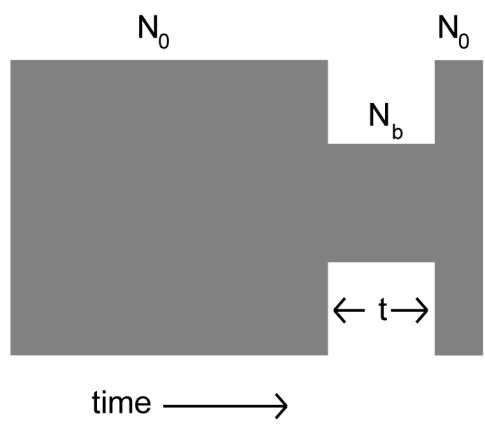